\documentclass[amsmath,amssymb,superscriptaddress,reprint,aip]{revtex4-2}
\usepackage{graphicx}
\usepackage{comment,xcolor}
\usepackage{siunitx}
\usepackage{bm}
\usepackage{braket}
\usepackage{comment}

\makeatletter
\let\@fnsymbol\@fnsymbol@latex
\@booleanfalse\altaffilletter@sw
\makeatother

\begin{document}
\title{Isotropic gap formation, localization, and waveguiding in mesoscale Yukawa-potential amorphous structures}

\author{Murat Can Sarihan}
\email[Corresponding author: ]{mcansarihan@ucla.edu}
\affiliation{Quantum Devices and Nanophotonics Research Laboratory, Department of Electrical and Electronics Engineering, Middle East Technical University, Ankara, Turkey}%
\affiliation{Mesoscopic Optics and Quantum Electronics Laboratory, Department of Electrical and Computer Engineering, University of California Los Angeles, Los Angeles, California, USA}%
\author{Alperen Govdeli}
\affiliation{Quantum Devices and Nanophotonics Research Laboratory, Department of Electrical and Electronics Engineering, Middle East Technical University, Ankara, Turkey}
\author{Zhihao Lan}
\affiliation{Department of Electronic and Electrical Engineering, University College London, Torrington Place, London WC1E 7JE, UK}
\author{Yildirim Batuhan Yilmaz}
\affiliation{Quantum Devices and Nanophotonics Research Laboratory, Department of Electrical and Electronics Engineering, Middle East Technical University, Ankara, Turkey}
\author{Mertcan Erdil}
\affiliation{Quantum Devices and Nanophotonics Research Laboratory, Department of Electrical and Electronics Engineering, Middle East Technical University, Ankara, Turkey}
\author{Yupei Wang}
\affiliation{Department of Electronic and Electrical Engineering, University College London, Torrington Place, London WC1E 7JE, UK}
\author{Mehmet Sirin Aras}
\affiliation{Mesoscopic Optics and Quantum Electronics Laboratory, Department of Electrical and Computer Engineering, University of California Los Angeles, Los Angeles, California, USA}
\author{Cenk Yanik}
\affiliation{Sabanci University Nanotechnology Research and Application Center, Sabanci University, Istanbul, Turkey}
\author{Nicolae Coriolan Panoiu}
\email[Corresponding author: ]{n.panoiu@ucl.ac.uk}
\affiliation{Department of Electronic and Electrical Engineering, University College London, Torrington Place, London WC1E 7JE, UK}
\author{Chee Wei Wong}
\email[Corresponding author: ]{cheewei.wong@ucla.edu}
\affiliation{Mesoscopic Optics and Quantum Electronics Laboratory, Department of Electrical and Computer Engineering, University of California Los Angeles, Los Angeles, California, USA}
\author{Serdar Kocaman}
\email[Corresponding author: ]{skocaman@metu.edu.tr}
\affiliation{Quantum Devices and Nanophotonics Research Laboratory, Department of Electrical and Electronics Engineering, Middle East Technical University, Ankara, Turkey}

\date{\today}

\begin{abstract}
\textbf{Abstract}

\textbf{Amorphous photonic structures are mesoscopic optical structures described by electrical permittivity distributions with underlying spatial randomness. They offer a unique platform for studying a broad set of electromagnetic phenomena, including transverse Anderson localization, enhanced wave transport, and suppressed diffusion in random media. Despite this, at a more practical level, there is insufficient work on both understanding the nature of optical transport and the conditions conducive to vector-wave localization in these planar structures, as well as their potential applications to photonic nanodevices. In this study, we fill this gap by investigating experimentally and theoretically the characteristics of optical transport in a class of amorphous photonic structures and by demonstrating their use to some basic waveguiding nanostructures. We demonstrate that these 2-D structures have unique isotropic and asymmetric band gaps for in-plane propagation, controlled from first principles by varying the scattering strength and whose properties are elucidated by establishing an analogy between photon and carrier transport in amorphous semiconductors. We further observe Urbach band tails in these random structures and uncover their relation to frequency- and disorder-dependent Anderson-like localized modes through the modified Ioffe-Regel criterion and their mean free path - localization length character. Finally, we illustrate that our amorphous structures can serve as a versatile platform in which photonic devices such as disorder-localized waveguides can be readily implemented.}
\end{abstract}

\maketitle

\section{Introduction}

A longstanding view in photonics has been that disorder is an undesirable feature, primarily due to its detrimental effect of light scattering with increased optical losses. However, recently, an alternative view has been emerging, namely that the behavior of light in optical structures defined by random permittivity distributions could exhibit a variety of fascinating properties and thus offer new opportunities in photonics for fundamental and applied research \cite{WiersmaNatPhot13,review_disorder21,WiersmaNature97}. For example, the existence of photonic band gaps in such disordered structures is somewhat counterintuitive, as the well-established paradigm had been that long-range periodic order was a prerequisite for the formation of band gaps via coherent wave scattering. Compared to the band gaps in photonic crystals (PhCs), which are highly anisotropic, band gaps in amorphous photonic structures are inherently isotropic and less sensitive to fabrication imperfections, thus providing improved functionality. Such isotropic band gaps of amorphous structures have been examined in the microwave \cite{JinPRB01,Mana13PNAS} and visible wavelengths \cite{RechtsmanPRL11}.

Amorphous photonic structures with artificially engineered disorder, characterized by random distributions of the refractive index, represent ideal platforms to study Anderson localization and related phenomena, such as enhanced or arrested wave transport \cite{segevScience17,RaedtPRL89,Schwartz07Nature,RiboliOL11,SegevNatPhot13,YangPRA10amorph,ConleyPRL14,HsiehNatPhys15, RoseOE13,Aubry20PRL,vynckarxiv21,segevPRB19, wiersmaNat08,segevSci11,segevNat20,segevNP12}. The origins of wave transport and the presence of Anderson-like localization in 2D amorphous photonic structures are highly debated topics \cite{vynckarxiv21,monsarratPRR2022}. The role of short-range order in band gap formation is well-established \cite{JinPRB01,ScheffoldPRL2016}, and the formation of band gaps under only short-range order is experimentally demonstrated in this article. Furthermore, a genuine Anderson localization transition was thought to occur only in 3D amorphous structures \cite{caoarxiv2022, ScheffoldNatComm2020}, while only infinite 2D structures can exhibit Anderson localization. This idea was challenged for out-of-plane (TM) polarization, where the scatterers behave as aligned in-plane monopoles, reducing the problem to scalar wave localization, for which Anderson transitions were observed in several studies. \cite{monsarratPRR2022,ConleyPRL14,RechtsmanPRL11, ScheffoldNatComm2020}. For in-plane (TE) polarization, each scatterer acts like a dipole with a different orientation, which is essentially a vector-wave scattering phenomenon long eluded the observation of localized states \cite{vynckarxiv21,perezarxiv2022,bachelardPRA2015, skipetrovPRB2021}. Finally, in a recent theoretical study, it was observed the presence of Anderson-like localization for TE (vector) waves in stealthy hyperuniform structures for a moderate density of scatterers. It was further demonstrated that hyperuniformity is not a prerequisite, the correlated disorder being sufficient to induce a transition to Anderson localization \cite{monsarratPRR2022, perezarxiv2022,torquatoPNAS2022}. Here, we looked for the signatures of TE (vector) wave localization in Yukawa-potential amorphous structures. 

Equally important, amorphous photonic structures have found technological applications to spectroscopy \cite{Redding13NatPhoton}, random lasers \cite{Wiersma13NatPhys,wiersmaNP17}, enhanced light-matter interactions \cite{Sapienza10Science,vynckNM12,segevPRL18}, and topological photonics \cite{szameitNat18,zhang20LSA,zhangPRB19,bourneJPA18,rechtsmanNature13}. Furthermore, amorphous photonic structures could also provide new or improved functionalities to photonic integrated circuits, wherein their inherent randomness renders immaterial the influence of fabrication imperfections \cite{MiyazakiPRB03,SarihanOCM19}.

In this study, we examine first the formation and dependencies of band gaps in Yukawa potential amorphous structures, along with wave localization, modified Ioffe-Regel factors, and applications to photonic integrated circuits. In measurements and theory, we implement Yukawa potential amorphous structures (YPAS) with short-range order, known to have spatially homogeneous and isotropic frequency dispersion characteristics. We nanofabricated and characterized these amorphous photonic structures in slab and waveguide configurations in the near-infrared regime, and uncovered the band gaps and key wave transport character dependencies on frequency and disorder strength. Furthermore, we uncover similarities between the properties of band tails of electronic and photonic amorphous structures. We also examine the nature of in-plane-polarization (TE) mid-gap states, namely, we seek to clarify whether they are Anderson-like localized optical modes that satisfy the Ioffe-Regel criterion. The results presented here can provide general guidelines for designing photonic components based on amorphous structures, thus forming the basis for designing key building blocks such as beam splitters, mirrors, optical cavities, and random lasers with isotropic emission.

\section{Results and Discussion}
\begin{figure}
\includegraphics[width=\columnwidth]{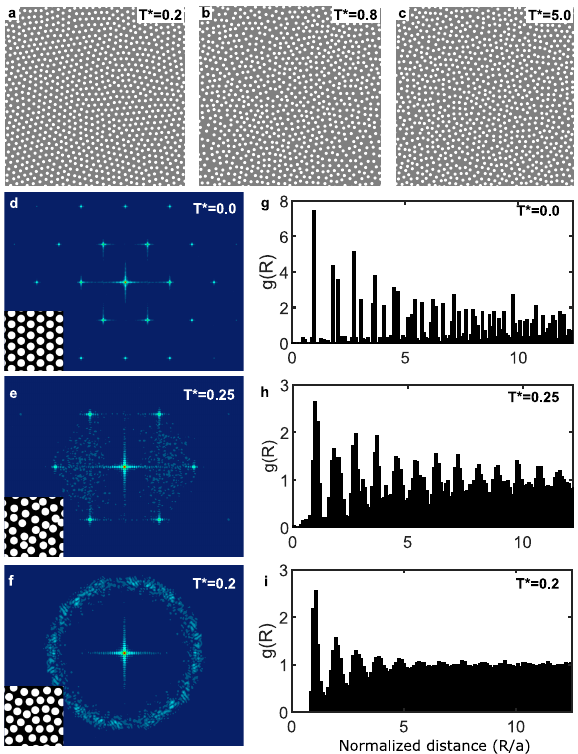}
\caption{{\bf Comparison of periodic and disordered lattices with Yukawa-potential amorphous structures}.(a), (b), (c) The permittivity distributions, $\epsilon(x,y)$, correspond to amorphous structures with increasing randomness. The randomness is quantified by the normalized temperature $T^*$, which is given as (a) 0.2, (b) 0.8, and (c) 5.0. (d), (e), (f) Fourier transform of the periodic, disordered, and amorphous structures, respectively. $\epsilon(x,y)$ are shown in the insets with the silicon (etched air hole) regions depicted in black (white). The periodic photonic crystal has a hexagonal symmetry with lattice constant $a$, whereas the disordered structure is generated from the periodic lattice by randomly perturbing each hole (see text for details). For the amorphous structure, the hole distribution is generated using a Metropolis Monte Carlo sampling approach. The normalized temperature, $T^*$, in the three cases, has values of 0, 0.25, and 0.2, respectively. The value of (e) indicates the range of the added randomness to (d) in terms of $\pm 0.25*a$ introduced to show the difference between disordered and amorphous structures.  (g), (h), (i) Corresponding radial distribution $g(R)$ of the air holes for the three lattice structures analyzed in (d), (e), and (f), respectively.} \label{fig:fig1}
\end{figure}

\textit{Yukawa-potential amorphous photonic structures.---}The electrical permittivity distribution of the YPAS is generated using a Metropolis Monte Carlo algorithm sampling all configurations to construct a Markov chain for transition probabilities of each particle (hole) \cite{RechtsmanPRL11,FrenkelSmit01,HynninenPRE03}. Consequently, the amorphous structure mimics the natural crystalline-to-glass transition of a semiconductor crystal, whereby each hole represents a point scattering particle in a colloidal, liquid-like structure, and the repulsive Yukawa potential governs the interaction between holes, $v(r)=(v_0/r)\exp(-r/r_{s}), r<r_0$ \cite{RechtsmanPRL11,HynninenPRE03}. Here, $r_{s}$ is the screening length, and $v_0$ is a constant characterizing the strength of the potential. A critical parameter of the generation algorithm is the normalized temperature, $T^*=k_\mathrm{B}T/v_0$, which quantifies the degree of randomness of a configuration. These parameters determine the Yukawa-potential equilibrium configuration, whose spatial characteristics are quantified by the average distance between the scatterers \cite{SarihanThesis18}, normalized in this work to the lattice constant of a periodic hexagonal lattice. In Figs.~\ref{fig:fig1}(a), \ref{fig:fig1}(b) and \ref{fig:fig1}(c), an example of hole distributions for $T^*=0.2$, $T^*=0.8$ and $T^*=5.0$ is given to emphasize the change in the equilibrium from low to high randomness. 

To illustrate the unique features of YPASs, we compare in Fig.~\ref{fig:fig1} the structural properties of periodic, disordered, and YPAS, quantified in both real and reciprocal spaces. The periodic structure is a hexagonal lattice with lattice constant, $a$, whereas the disordered structure is constructed from the periodic lattice by displacing each hole along the $x$- or $y$-axes by a quantity $\delta$ obeying a uniform random distribution defined in the interval, $[-0.25a/2,0.25a/2]$ marked with a pseudo-temperature parameter 0.25 in Fig.~\ref{fig:fig1}(e). The Yukawa-potential configurations are different from the periodic and disordered structures in several key aspects, which is apparent by performing a Fourier transform analysis of the resulting hole distributions \cite{HsiehNatPhys15,WiersmaNatPhot13,GaoSciRep13,RiboliNatMater14, rojasPRL2004}. The Fourier transform of periodic (Fig.~\ref{fig:fig1}(d)) and disordered (Fig.~\ref{fig:fig1}(e)) permittivity distributions have Bragg peaks, which is a signature of periodicity and long-range order. We note that, for the disordered structure, the added perturbation smears out and broadens the Bragg peaks, but their occurrence signifies that the underlying structure periodicity is preserved. The Fourier transform of the YPAS (Fig.~\ref{fig:fig1}(f)) noticeably, however, is isotropic and shows no local maxima.

The range of spatial order is characterized by the radial distribution function, $g(R)$, which is equal to the number $dN$ of holes located in the annulus $(R, R+dR)$ \cite{liu2018}. Specifically, as per Fig.~\ref{fig:fig1}(g),  $g(R)$ shows discrete peaks even at large distances for the hexagonal lattice, confirming the long-range order. Upon adding disorder, these peaks begin to smear out, yet the underlying pattern is still preserved up to 10 unit distances, as per Fig.~\ref{fig:fig1}(h). In the amorphous structure, however, the peaks of the radial distribution function are attenuated and merged with the pedestal at about $5$ unit distances (Fig.~\ref{fig:fig1}(i)), indicating the suppression of long-range order.

\begin{figure}[t]
\includegraphics[width=\columnwidth]{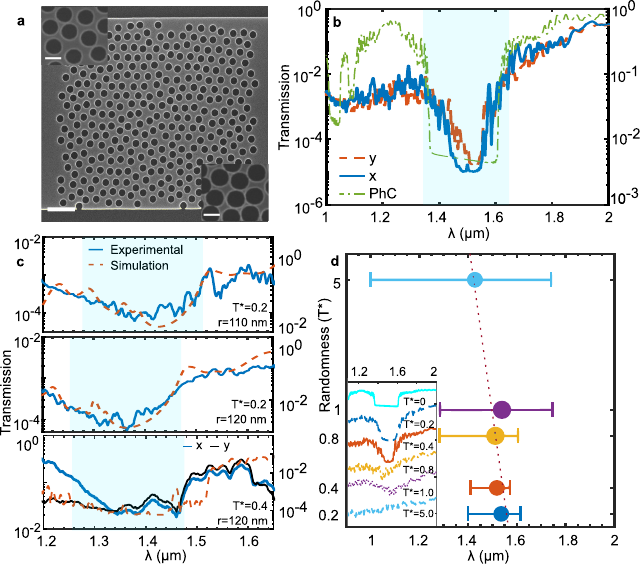}
\caption{{\bf Photonic band gaps in Yukawa-potential amorphous structures}. (a) Scanning electron micrograph (SEM) of an amorphous nanostructure with $T^*=0.2$, $r=\SI{110}{\nm}$, and $h=\SI{220}{\nm}$, fabricated onto a silicon-on-insulator (SOI) wafer. Scale bar: \SI{1}{\micro\meter}. Insets show zoom-in SEMs of two structures with $r=\SI{110}{\nm}$ (top left) and $r=\SI{120}{\nm}$ (bottom right). Scale bars: \SI{250}{\nm}.  (b) Comparison of transmission spectra of the PhC and amorphous structure with the same thickness and hole radius and with $T^* = 0.4$. The latter are calculated for propagation along the $x$- and $y$-axes. The right y-axis indicates the transmission in arbitrary units for the PhC band gap (green) scaled to show the match between amorphous and PhC band gaps. The cyan shade is a marker for the band gap region of the reference photonic crystal. (c) Measured transmission spectra corresponding to YPAS with $T^*=0.2$ and hole radius of top: \SI{110}{\nm} and middle: \SI{120}{\nm}, and with $T^*=0.4$ and bottom: \SI{120}{\nm} respectively.  The blue and black lines are the experimentally measured spectra in $x$- and $y$- directions and smoothed via Savitsky-Golay filtering; the red line is the numerically computed spectra by importing the actual fabricated structure in the software. The cyan shade is a marker for the band gap region. (d) Computed band gap dependence on $T^*$, where the line bars indicate the region of suppressed transmission according to 10-dB fall-off from the dielectric and air band lev-
els, while the circle pointer shows the minimum transmission point. The dotted line is a guide to the eye. In the inset, the transmission spectra determined for different values of $T^*$ are shown by stacking each line, whose y-axis values correspond to transmission in arbitrary values, with the same scale for each line. The PhC slab corresponds to $T^*=0$ and for comparison purposes, the spectrum in this case and in (b) is determined by averaging the PhC spectra over the propagation directions.} \label{fig:fig2}
\end{figure}

 \textit{Isotropic band gap in the amorphous structures.---}Previous studies at microwave frequencies showed that PhCs and amorphous structures with the same characteristic length share the first band gap because in both cases, the band gap originates from short-range order\textcolor{black}{ \cite{JinPRB01, ScheffoldPRL2016}.} Thus, we considered a reference PhC slab with a hexagonal lattice of air holes. Furthermore, for all three configurations (PhC, disordered, and amorphous structures), the characteristic length, $a$ (i.e., the averaged nearest-neighbor distance between holes), slab thickness, $h$, and hole radius, $r$, are kept the same, so that in all three cases there is a common transverse-electric (TE) band gap. Here, TE mode designation is in accordance with photonic crystal slab terminology, where even modes are called TE-like and odd modes are called TM-like. According to our simulations, it is also possible to form TM (transverse-magnetic) band gap when dielectric rods are used inversely. For the parameters used in our experiments, namely $a=\SI{400}{\nm}$, $h=\SI{220}{\nm}$, and $r=\SI{110}{\nm}$, the mid-gap wavelength is about \SI{1500}{\nm}. Figure \ref{fig:fig2}(a) shows the scanning electron micrograph (SEM) of the amorphous structure with \SI{400}{\nm} average distance between holes, fabricated via electron-beam nanolithography followed by inductively coupled plasma reactive ion etching on silicon-on-insulator (SOI) wafers with \SI{220}{\nm} thickness of the silicon device layer. Each sample is suspended to form an air-bridge membrane with buffered oxide etch (BOE) and thus creates a symmetric planar membrane configuration. 

The dispersive properties of our structures are investigated using finite-difference time-domain computations \cite{JohnsonOE01,OskooiCPC10MEEP}. Fig.~\ref{fig:fig2}(b) shows the computed transmission spectra of the PhC and amorphous structures. These spectra show that for the YPAS with $T^* = 0.4$ and $r=\SI{110}{\nm}$, the optical transmission is mainly independent of the propagation direction, the mobility edges fit for both \textit{x} (blue) and \textit{y} (red) directions, which matches with the overlapping photonic band gap of the PhC (green) illustrating that the structure is isotropic. In contrast, the PhC structure is highly anisotropic, with the transmission different along the $\Gamma-M$ and $\Gamma-K$ symmetry axes (detailed in Supplementary Note 1, Figure S2(b)). We note that, while the gap width is similar for both photonic structures, the isotropy provided by the YPAS can be employed to achieve all-angles broadband mirrors, devices with reduced angle sensitivity, and waveguiding along arbitrary directions.

For the $T^*=0.2$ realization, the transmission is measured for two samples with hole radii of \SI{110}{\nm} and \SI{120}{\nm}, whereas for the $T^*=0.4$ realization, transmission is measured for a sample with $r = \SI{120}{\nm}$ in both \textit{x-} and \textit{y-} directions. All three measurements are depicted in Fig.~\ref{fig:fig2}(c) with solid blue lines. All the measured spectra match well with our simulation results (dashed red), where we extracted the actual hole positions and average hole radius from the SEM images of the device. The band tails are fitted between mid-gap and dielectric and air band edges according to the Urbach expression, $T \approx e^{(\lambda-\lambda_\mathrm{midgap})/\Delta\lambda}$, where $\Delta\lambda$ is the effective width of the tail \cite{StudenyakIJOA14}. In this configuration, dielectric band tail widths are measured as 12.83 nm for $r = \SI{110}{\nm}$ and 14.98 nm for $r = \SI{120}{\nm}$, for $T^*=0.2$. For the air band tail, the change is more pronounced, 33.27 nm for $r = \SI{110}{\nm}$ and 39 nm for $r = \SI{120}{\nm}$, indicating the higher density of localized states at the air band with larger holes. The tail widths are longer than the widths fitted in the numerical spectra, which might be due to fabrication-related disorders. In addition, we note that for the larger hole radius of \SI{120}{\nm}, a slight blue-shift of the mid-gap wavelength is observed, which agrees with predictions based on the perturbation theory for a larger radius, where a larger amount of dielectric material is removed from the homogeneous slab. When $T^*$ increases from 0.2 to 0.4, the mid-gap wavelength remains largely unaffected (cf. Figs.~\ref{fig:fig2}(c)). Importantly, the spectra are asymmetric, displaying a steeper slope at the longer-wavelength edge. The origins of the longer-wavelength steeper slope and the mid-gap blue-shift are discussed together with the localization strength and air-dielectric band perturbation in the next subsection.

In Fig.~\ref{fig:fig2}(d), we examine the band gap widths for different values of the normalized temperature, $T^*$ (the degree of randomness increases with $T^*$), with the corresponding transmission spectra plotted in the inset. The mid-gap wavelengths are selected as the minimum transmission point, while the gap boundaries are set according to 10-dB fall-off from the dielectric and air band levels. We observe that the band gap width for $T^*=0.2$ and $T^*=0.4$ has similar values and is almost equal to that of the PhC ($T^*= 0$). At larger $T^*$, an increasing number of localized states emerges inside the band gap and enhances the light scattering. As a consequence, the transmission is suppressed almost entirely. \textcolor{black}{We further observe the localized-to-extended mode transition with increasing $T^*$, where the increase in localized mode density causes a reduction in the gap width of $T^*=0.4$, from 215 nm to 162 nm. However, at  $T^* > 0.8$, we observe a shallow gap-like regime where the transmission is suppressed only 10 dB compared to dielectric and air bands. This localized-to-extended transition will be examined further using effective modal width and localization length.  Moreover, we note that, with increasing $T^*$, the mid-gap wavelength is slightly blue-shifted within a region of 1.45 and 1.5 $\mu$m, due to the change in the lowest transmission point with localized-to-extended mode transition. The only noticeable exception is $T^* = 1.0$, where high statistical variations in the configuration change the lowest transmission point.} For our experimental measurements, we chose $T^* = 0.2$ and $T^* = 0.4$, with hole radii of $0.275a$ and $0.3a$ respectively. The nanofabrication feasibility of the structures and accessibility of probing the angular dependencies guided us in choosing these values.

\begin{figure}
\includegraphics[width=\columnwidth]{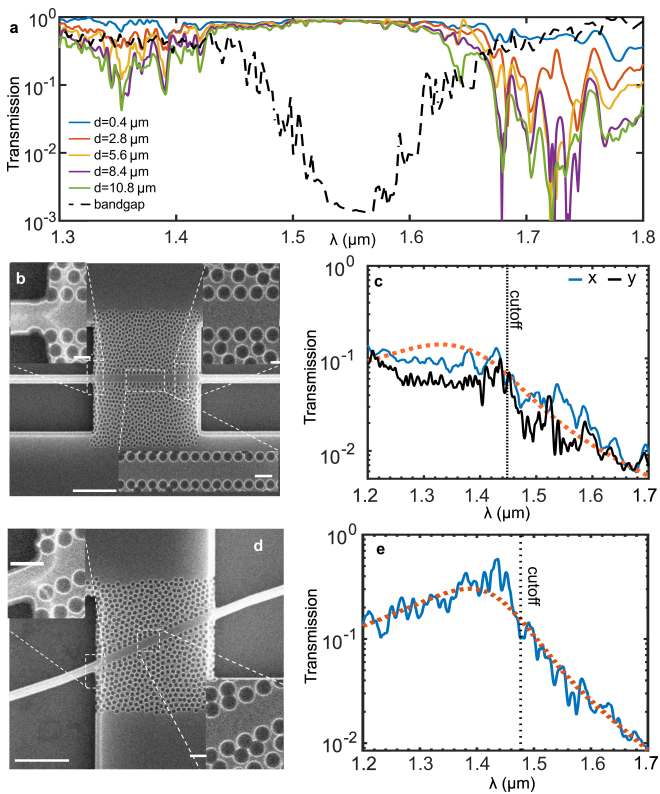}
\caption{{\bf Experimental characterization of the band gap of Yukawa-potential amorphous structures and corresponding waveguides}. (a) Transmission spectra of straight waveguides embedded in an amorphous structure with $w=\SI{901}{nm}$ and $r=\SI{100}{nm}$. The dashed lines indicate the transmission spectra through a section of amorphous structure, whereas the distance from the excitation source is indicated in the legends. (b) Scanning electron microscopy (SEM) of an optical waveguide based on a YPAS with $T^*=0.2$ and $r=\SI{120}{nm}$. The waveguide is oriented along the $x$-axis and has a width of \SI{901}{\nm}. Scale bar: \SI{5}{\micro\meter}. Insets show zoom-in SEMs of different regions of the waveguide. Scale bars: \SI{500}{\nm}. (c) Measured (blue and black lines) and fitted (red line) transmission spectra of the amorphous waveguide structure oriented along $x$-axis (blue) and $y$-axis (black). The \SI{3}{\decibel} cutoff wavelength is $\sim$\SI{1448}{\nm}. (d) The same as in (b), except that the waveguide is oriented at an angle of \SI{23}{\degree} with respect to the $x$-axis. (e) The same as in (c), but for the waveguide in (d). The \SI{3}{\decibel} cutoff wavelength is $\sim$\SI{1478}{\nm}.} \label{fig:fig3}
\end{figure}

\textit{Flexible waveguides with Yukawa-potential amorphous structures.---}Following the experimental observation of photonic band gaps and Anderson-like localization of YPAS, we demonstrate next that such disordered optical media can be employed to optical nanodevices, such as sub-wavelength waveguides. With the isotropy of the band gap of the amorphous structures, the orientation of the waveguide can be arbitrarily chosen. This is in stark contrast with the case of 1D waveguides in PhC slabs, for which the waveguide mode dispersion depends strongly on the waveguide-wavevector orientation. Likewise, more complex waveguide configurations, such as sharp bends and 1-to-$N$ splitters, can be readily implemented using our amorphous materials. Note that to ensure that the waveguide has a constant width, we added a row of equally spaced holes adjacently to both sides of the waveguide \cite{MiyazakiPRB03}, as per Fig.~\ref{fig:fig3}(b). Moreover, guided by our previous simulations \cite{SarihanOCM19}, the waveguide width, $w$, was optimized so that the maximum transmission is achieved at the mid-gap wavelength, a condition fulfilled for $w\approx\SI{901}{\nm}$ (more details on the waveguide design process are in our Supplementary Note 3). Fig.~\ref{fig:fig3}(a) shows the simulated transmission spectra of the optimized waveguide with different lengths, for
$r=\SI{100}{\nm}$. At the optimized mid-gap wavelength, the simulation indicates a \SI{0.5}{\decibel} insertion loss, without any noticeable propagation loss up to $\SI{10.8}{\um}$, a very promising result for chip-scale applications.

Figure \ref{fig:fig3}(c) shows the measured transmission spectrum for a waveguide oriented along the $x$- and $y$- axis. We note that the measured value of the \SI{3}{\decibel} cutoff wavelength agrees well with the simulation results presented in Fig.~\ref{fig:fig2}(c). To further validate our conclusions, we fabricated another waveguide in the same amorphous structure with the same geometrical parameters, except that it was oriented at an angle of \SI{23}{\degree} to the $x$-axis (see the SEM in Fig.~\ref{fig:fig3}(d)). The corresponding transmission spectrum is shown in Fig.~\ref{fig:fig3}(e) and is very similar to that presented in Fig.~\ref{fig:fig3}(c) in terms of cutoff wavelength and baseband propagation loss. These results further support the band gap isotropy of our YPAS, which is the indispensable condition for long-sought complete band gaps in planar structures.

\begin{figure}
\includegraphics[width=\columnwidth]{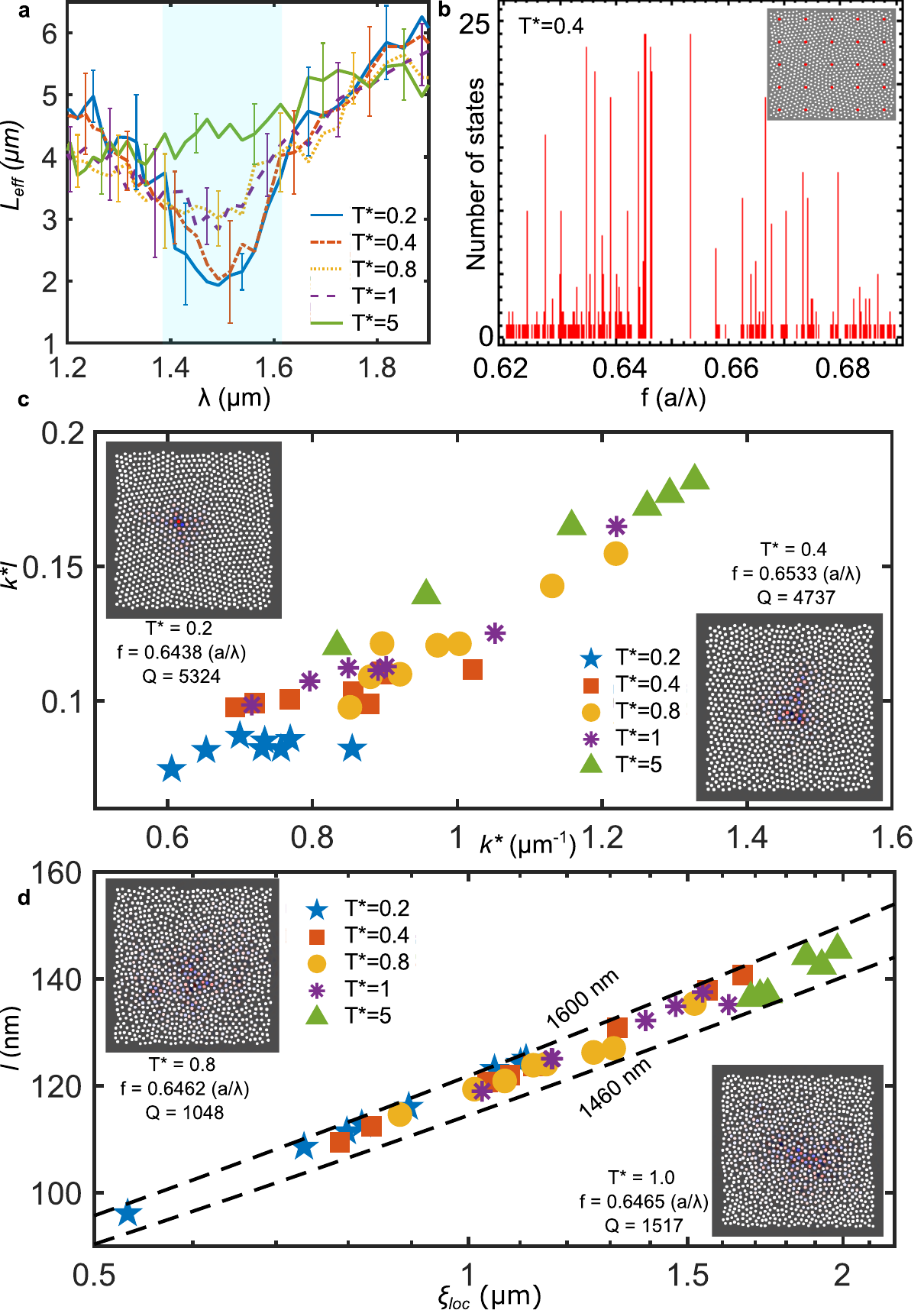}
\caption{{\bf Wave localization in Yukawa-potential amorphous structures}. (a) Wavelength dependence of the scattering length for different $T^*$. The error bars are the standard deviation of the effective length corresponding to a given $T^*$ and to a set of multiple realizations of disorder. (b) Number of modes at the normalized frequencies above but the nearby edge of the dielectric band (K-point), for $T^*=0.4$. The inset shows the modal probes used during calculations. (c) Modified Ioffe-Regel factor vs. the shifted wave number, $k^{*}$. (d) Mean free path vs. localization length, determined for the modes in (c). The dashed lines correspond to the wavelengths of the top edge of the dielectric band ($\lambda=\SI{1600}{\nm}$) and mid-gap ($\lambda=\SI{1460}{\nm}$).The insets of (c) and (d) show example modal distributions of the modes with the highest quality factor (localization) from $T^*=0.2$ to $T^*=5.0$. A full list of the selected modes is given in Supplementary Note 2, Table S1.} \label{fig:fig4}
\end{figure}

\textit{Anderson-like localization in Yukawa-potential amorphous structures.---} After investigating experimentally the formation of optically isotropic photonic band gaps in YPASs, we further analyzed the characteristics of optical transport in these structures theoretically. Of interest is the Anderson light localization in our YPASs. Particularly, the effective length of the local field distribution, $L_{\textrm{eff}}$, which quantifies the strength of the field localization and is defined as $L_{\textrm{eff}}\equiv P^{-1/2}=\{ [\int I(\mathbf{r})^2d\mathbf{r}]/[\int I(\mathbf{r})d\mathbf{r}]^2\}^{-1/2}$. Here, $I(\mathbf{r})$ is the spatial distribution of the optical field, $\mathbf{r}=(x,y)$ is the in-plane position vector, and $P$ is the inverse participation ratio \cite{Schwartz07Nature}. The effective length was calculated for different values of the temperature $T^*$, and a statistical average over 10 samples was performed for each temperature. The effective length, whose wavelength dependence is depicted in Fig.~\ref{fig:fig4}(a), varies linearly with the number of available states in the amorphous structure.\textcolor{black}{ The results in Fig.~\ref{fig:fig4}(a) show that inside the band gap $L_{\textrm{eff}}$ increases with $T^*$, with an averaged mid-gap value of $L_{\textrm{eff}}=\SI{1.93}{\micro\meter}\approx5a$ being observed for $T^{*}=0.2$. Moreover, for the same $T^{*}$, $L_{\textrm{eff}}$ decreases by about $3\times$ when the wavelength varies from the edge of the band gap to its center, underlying the transition of these edge modes away from weak localization and diffusive transport.}

As can be seen in Fig.~\ref{fig:fig4}(a), $L_{\textrm{eff}}$ is smaller and more dispersive at the air-band edge (short-wavelength edge) of the band gap as compared to its behavior at the dielectric-band edge (long-wavelength edge). That is, there is spectral asymmetry with respect to the mid-gap, as noted earlier. A similar phenomenon is observed in condensed matter physics, where the transition of a liquefied crystalline semiconductor to a glassy medium is accompanied by the formation of what is known as Urbach tails \cite{FeddersPRB98, StudenyakIJOA14}. They emerge when the valence and conduction bands begin to extend into the band gap due to the creation of localized trap states near the band edges. These exponential tails are revealed by the absorption spectra and are primarily determined by the thermal motion inside the lattice, disorder in the crystal structure, and impurity atoms \cite{DraboldPRB11}. From the tight-binding model, the valence (conduction) band is of bonding (antibonding) nature \cite{FeddersPRB98}. Moreover, the valence (conduction) tail states are associated with short (long) bonds inside the lattice, and since the short bonds are affected more by disorder, the corresponding states are more localized, and the valence band tail is broader \cite{DraboldPRB11}. In other words, the Urbach tails of the valence and conduction bands have different slopes; namely, the slope of the valence band tail is smaller than that of the conduction band tail \cite{FeddersPRB98}. This behavior is observed in our Yukawa-potential material, which can be viewed as a photonic analog of amorphous material in condensed matter physics.

As the plots presented in Fig.~\ref{fig:fig2}(b) and Fig.~\ref{fig:fig4}(a) suggest, the wavelength dependence of $E$-field localization within the air and dielectric band regions has different slopes. Specifically, for $T^*=0.4$ case shown in Fig.~\ref{fig:fig2}(b), the $1/e$ intensity fall-off length for the air band tail is \SI{16.54}{\nm}, whereas a steeper fall-off of \SI{9.03}{\nm} is observed for the dielectric band tail. Hence, the air band is more sensitive to perturbations, a result that can be explained as follows: for the air band, a more significant fraction of the $E$-field is located in the hole regions, and therefore, the perturbation associated with a random displacement of the location of the holes is more substantial in this case. As the wavelength increases across the band gap, the $E$-field is pushed into the dielectric regions. This behavior is similar to that seen in PhCs, where the difference between $E$-field localization corresponding to two consecutive bands leads to a band gap formation \cite{JoannopoulosBook08PhC}. Moreover, as expected, the wavelength dependence of the $E$-field localization levels off as $T^*$ increases. Note that the wavelength dependence of the $H$-field localization shows a behavior opposite to that of the $E$-field (detailed in Supplementary Note 1). In addition, this perturbation affects the modes in the air band more, leading to a more extensive spectral broadening of these modes. As a result, the air band tail would have a smaller slope as compared to that of the dielectric band, a conclusion fully supported by the transmission spectra in the subsequent Figs.~\ref{fig:fig2}(b) through \ref{fig:fig2}(d) \cite{Schwartz07Nature}.

Now, let us examine the spatial distribution of the modes located inside the band gap and quantify their degree of localization. We first compute the optical modes of amorphous structures characterized by each $T^*$, and which are located in the region of the dielectric-band tail, i.e., close to the $K$-symmetry point. An example sampling of such modes is illustrated in Fig.~\ref{fig:fig4}(b) (detailed in Supplementary Note 2, Figures S5 and S6. From this set of modes, we select those with large $Q$-factor, which are presumably more localized (see the insets of  Figs.~\ref{fig:fig4}(c) and \ref{fig:fig4}(d)) and thus are not affected by the structure boundaries. Since we want to characterize structures ranging from weakly perturbed PhCs ($T^{*}\ll1$) to YPAS with a large $T^{*}$, we used the modified Ioffe-Regel factor, $k^*l$ \cite{JohnPhysToday91}, where $k^*=|k-k^K_{\textrm{be}}|$. Here, $k$ is the wavenumber, $k^K_{\textrm{be}}$ is the wavenumber of the dielectric-band edge (the $K$-point), and $l$ is the mean free path. The modified Ioffe-Regel factor implicitly assumes the existence of a band gap, which strongly affects the free-photon density of states, while the scatterers are much smaller than the operating wavelength. Under these circumstances, the condition for the Anderson-like localization is $k^*l\simeq1$.

The Ioffe-Regel factor of the selected modes, determined for different values of $T^{*}$, is presented in Fig.~\ref{fig:fig4}(c). These plots reveal that the Ioffe-Regel factor is well below 1, indicating that the modes are truly localized Anderson modes. Moreover, the mode localization length, $\xi_{\textrm{loc}}$, can be determined by fitting the mode intensity profile to an exponential function, $I~\sim \exp(-2r/\xi_{\textrm{loc}})$ \cite{Schwartz07Nature, LeeSciAdv18}. Consequently, the mean free path, $l$, can be obtained from the fitted parameter by the definition, $\xi_{\textrm{loc}}= l \textrm{exp}(\pi k l /2)$, such that the modes in Fig.~\ref{fig:fig4}(c) can be represented in the $\xi_{\textrm{loc}}-l$ plane, as per Fig.~\ref{fig:fig4}(d).  From this figure, one can see that $\xi_{\textrm{loc}}$ increases with $T^{*}$, with $l$ ranging from \SIrange{100}{150}{\nm}. This is only a fraction of the lattice constant, $a$, whereas $\xi_{\textrm{loc}}\approx\SI{1}{\mu m}=2.5a$. This demonstrates that these states are indeed localized at the scale of the lattice constant. Note that these conclusions hold for the modes in the air-band tail, too. 

Amorphous photonic structures have been largely overlooked when seeking to develop new photonic technologies, chiefly due to their complex fabrication and the statistical rather than deterministic nature of their properties. Contrary to this commonly adopted viewpoint, we demonstrate that amorphous photonic structures can be employed as a photonic material platform onto which planar photonic integrated circuits with complex configurations can be readily implemented. The key ingredient that makes this possible is the existence of an isotropic band gap originating from short-range order. Importantly, from a practical standpoint, the optical properties of these amorphous structures can be easily tuned by varying the two parameters that control their structure, namely the average inter-hole distance and effective temperature. This suggests that the amorphous photonic material platform introduced here can be used to implement key photonic devices, including Mach-Zehnder interferometers, multiplexers, and optical microcavities, and represents a practical testbed for new ideas emerging in the field of wave dynamics in random media.

\section{Methods}

\textit{Generation of amorphous photonic structure configurations.---} To investigate amorphous structures computationally, we employed a Monte Carlo algorithm for the generation of permittivity distributions that do not contain long-range order. For two-dimensional (2D) photonic crystal (PhC) slabs, one approximates rod or hole scatterers with hard disks or point particles located in the transverse $(x-y)$ plane, while along the $z$-axis, one imposes an index guiding condition. The flow of the Monte Carlo algorithm used to generate such amorphous photonic configurations is described below \cite{FrenkelSmit01,SarihanThesis18}.

\textit{A.} An input configuration is generated. In particular, we used as a starting structure a PhC slab configuration with a hexagonal lattice of air holes in a silicon slab. Using an ordered crystal lattice as a starting point allows one to use it as a reference periodic structure when investigating the transport properties of the resulting amorphous structure. Alternatively, a previously generated amorphous configuration can be used as a starting distribution, too, thus reaching equilibrium more rapidly. The potential energy of the initial configuration is calculated assuming that the inter-particle interactions in the system are described by a Yukawa-potential. Statistically, rules of canonical ensembles (constant number of elements $N$, volume $V$, and temperature $T$) are used.

\textit{B.} A random displacement is applied to a scatterer, randomly selected from the distribution. The maximum allowed value of the displacement is adjusted so that it results in an acceptance displacements ratio, $\rho$, in the range of \SIrange{40}{60}{\percent}. The difference in the potential energy of the configuration of scatterers, induced by this displacement, is then computed.

\textit{C.} The acceptance ratio of displacements is determined using the probability calculated from the Metropolis criterion:
\begin{align}
\rho_{i\rightarrow f} = \min\left(1,e^{-\frac{U_{f}-U_{i}}{k_\mathrm{B} T}}\right),
\end{align}
where $U_{f}$ and $U_{i}$ are the system energies in the final ($f$) and initial ($i$) state, respectively.

\textit{D.} The procedures described in steps $B$ and $C$ are repeated for different scatterers until equilibrium is reached. Specifically, the equilibrium is reached when the system energy converges to a steady-state value, that is when the system energy no longer changes upon randomly displacing a scatterer.

\textit{Numerical methods.---}The photonic band structure of the reference PhC slab, for both the transverse magnetic (TM) and transverse electric (TE) polarizations, have been numerically calculated using MPB (MIT Photonic Bands) software \cite{JohnsonOE01}. Moreover, transmission spectra and local density of states (LDoS) were computed using the finite-difference-time-domain (FDTD) method implemented in MEEP (MIT Electromagnetic Equation Propagation) software \cite{OskooiCPC10MEEP}. \textcolor{black}{Simulated transmission spectra are normalized using the output of a reference simulation of the same substrate without amorphous structure.} 

To study these mid-gap localized modes, we first probe them using the FDTD method implemented in MEEP \cite{OskooiCPC10MEEP}. Note that as these modes may be highly localized spatially, to effectively probe them, we set 25 probes covering the photonic structure uniformly, as illustrated in Fig.~\ref{fig:fig4}(b). As field excitation, we used a source whose spectrum was broad enough to entirely cover the band gap. We used the \textsf{Harminv} procedure available in MEEP to analyze the spectral characteristics (normalized frequency, $f$, and quality factor, $Q$) of the localized modes. For effective localization length calculations, the integrals are performed in the transverse plane crossing through the middle of the amorphous crystal slab.

\textit{Supercontinuum transmission measurements.---}Using a NKT Photonics SuperKCompact supercontinuum light source, we have conducted a broadband spectral analysis of the fabricated structures. The laser was coupled to the amorphous structures and waveguides through an adiabatically tapered waveguide-lensed fiber system. The output is collected by a similar tapered waveguide-lensed fiber system. We analyzed the spectra collected by the fiber with a Thorlabs optical spectrum analyzer, whereas an InGaAs camera was used to monitor the optical coupling to the on-chip waveguides. For reference, the measured spectra were normalized to the output of channel waveguides with uniform cross-sections. Experimental data shown in Figures \ref{fig:fig2} and \ref{fig:fig3} are given as normalized transmission ratio with the aforementioned method.

\textbf{Data Availability.---}
The data that support the findings of this study are available from the authors upon reasonable request.

\textbf{Code Availability.---}
The underlying code for this study is not publicly available but may be made available to qualified researchers upon reasonable request from the corresponding author.

\textbf{Acknowledgements.---}This work was supported by the Scientific and Technological Research Council of Turkey (TUBITAK), Grant No: 117E178, by the European Research Council (ERC; Grant No. ERC-2014-CoG-648328), and by the Office of Naval Research in the United States (N00014-21-1-2259). The authors also acknowledge discussions with Aaswath Raman, Sakir Erkoc, and Mikael C. Rechtsman. 

\textbf{Author contributions.---}
M.C.S., C.W.W, and S.K. designed and led the project. M.C.S. and Y.B.Y. designed the waveguide structures. M.C.S., Z. L., and Y. W. conducted numerical simulations and spectral analysis. M.S.A. and C.Y. performed the device nanofabrication. M.C.S., A.G., and M.E. conducted the measurements.   M.C.S., N.C.P., C.W.W., and S.K wrote the manuscript, with contributions from all authors. 

\textbf{Additional information.---}
The authors declare no competing interests. Reprints and permission information are available online at http://www.nature.com/reprints/. Correspondence and material requests should be addressed to M.C.S, N.C.P, C.W.W, and S.K.
\newpage

\end{document}